\documentclass[12pt]{iopart}

\usepackage{graphicx}

\begin{document}

\title[]{Sub shot-noise frequency estimation with bounded \textit{a priori} knowledge}

\author{Changhun Oh and Wonmin Son\footnote{Corresponding author: sonwm71@sogang.ac.kr}}

\address{Department of Physics, Sogang University, Mapo-gu, Shinsu-dong, Seoul 121-742, Korea}

\ead{sonwm71@sogang.ac.kr}

\begin{abstract}
We analyze an efficient frequency estimation scheme that is applied to measure the unknown frequency of an atomic state in Ramsey spectroscopy. The scheme is employing appropriate combinations of uncorrelated probe atoms and Greenburgur-Horne-Zeilinger (GHZ) type correlated probe atoms to estimate its frequency. The estimation value of frequency is obtained through the Bayesian analysis of the final measurement outcomes. The proposed scheme allows us to obtain better precision than the scheme without quantum correlation and it also prevents us from ambiguity in the frequency estimation procedure with GHZ correlations only. We show that the scheme can beat the shot-noise limit and, in addition, it is found that there is the trade-off relation between the precision of the frequency estimation and the decoherence rate in the atomic states.
\end{abstract}

\vspace{2pc}
\noindent{\it Keywords}:

\maketitle

\section{Introduction}

Measurement of a physical system with an arbitrary precision is an important task in many fields of experimental physics and technologies. The ultimate goal of the precision measurement is to attain the highest accuracy when a physical system is measured at a cost of given resources. In a real situation, the aim is needed to be achieved under the constraint of limited resources such as the finite number of trials and the limited duration time of the experiment. This situation is also true in the case of Ramsey spectroscopy experiment which is a scheme to measure the dynamical evolution of an atomic state and estimate the resonance frequency of the atomic state  \cite{Ramsey}. Although precision in the measurements can be improved with the number of trials, it is also true that the limited resource restricts the possible number of repetitions of the measurement. Therefore, it is required to devise a method how to utilize the given resources as efficient as possible.

A possible method to achieve a higher accuracy in the precision measurement is to employ quantum entanglement. The examples are to use coherent-squeezed state in gravitational wave detector \cite{caves}, number entangled states of photons (so-called NOON states) in Mach-Zehnder interferometer \cite{noon} and GHZ states of atoms in frequency estimation with Ramsey spectroscopy \cite{wineland1994, wineland1996}. In contrast to the conventional measurement scheme that allows us to attain the shot-noise limit only, $\delta\theta\sim 1/\sqrt{N}$ \cite{Itano 1993}, the scheme with entangled states helps us to obtain the Heisenberg limit, $\delta\theta\sim 1/N$ \cite{giovannetti2006,maccone,davidovich2011} where $\theta$ is an unknown parameter to be measured and $N$ is the number of employed resources. However, the precision measurement scheme is sometimes restricted if a prior knowledge about the value of a parameter to be measured is not provided. In the case of phase estimations, there is recent work about an estimation scheme that is performed by using entangled coherent states without a prior knowledge about an unknown phase \cite{ecs}. The reason that the precise estimation of a completely unknown phase is possible is that without any prior knowledge about the unknown phase, we know that phases lie in $[0,2\pi)$. In general, the precision measurement scheme is only useful when we try to achieve a better precision on vaguely known values. If we do not have any knowledge on the measurement values, the scheme does not work well.

Especially, when there is periodicity of probabilities in each measurement, the measurement data cannot be used to specify the single value of phase estimation and results in more than one estimation value over the period. Thus, if we do not know the range of the parameter sharp enough, it is not possible to estimate the precise value of its phase and leaves ambiguity. Since the period of probabilities for the measurement with entangled states is even shorter than that of unentangled states, entanglement-based measurements give rise to a more serious ambiguity problem in the determination of an estimation value. It can be shown that in the scenario having a prior knowledge on the periodicity, the measurement scheme with highly entangled states merely attains an equal precision to that with uncorrelated states in the frequency estimation. Therefore, there will be no benefit of using entangled states in the case with limited {\it a priori} knowledge provided. Here, we propose measurement schemes that allow us to overcome the ambiguity of $2\pi$ periodicity and, at the same time, obtain a better precision beyond the shot-noise limit using quantum entanglement.

There have been prior discussions that, when highly entangled states are used, periodicity of probabilities in each measurement outcome sometimes prevents us from determining a unique estimation value and it can be overcome in idealized cases \cite{pezze2007, higgins, dunningham}. In the works, two different settings are considered (i) using GHZ states with particle numbers $1,2,4,...,2^{p-1}$, and (ii) using combinations of an uncorrelated state and GHZ states in order to avoid ambiguity. Each scheme has been discussed in interferometric systems for phase estimation and the introduced decoherence model was photon losses. In this paper, we consider more general situation with a different decoherence effect that occurs in trapped ions. We compare the results of the two schemes in the absence of decoherence and in the presence of decoherence, and conclude that while the first scheme allows us to achieve the Heisenberg limit in the ideal case, the second scheme becomes more advantageous as the decoherence rate increases and attains a sub shot-noise precision.

We organize this article in the following. In section 2, we review frequency estimation procedure in Ramsey spectroscopy when the atoms are uncorrelated and correlated under decoherence effects. Furthermore, we bring up an ambiguity problem that occurs in the frequency estimation. In section 3, we analyze two different schemes that overcome the ambiguity problem and, at the same time, improve precision by using entanglement, and compare the attainable precision of the proposed schemes with that of the conventional scheme using no entanglement. Finally, we summarize our works in section 4.

\section{Ramsey Spectroscopy}
\subsection{Standard frequency estimation scheme under decoherence}
\begin{figure}
\center
\includegraphics[scale=0.7]{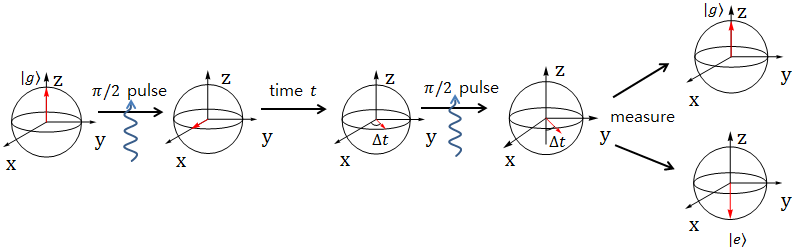}
\caption{The operation of Ramsey spectroscopy using a single ion is shown. The probability of detecting $|g\rangle$ is $\sin^2(\Delta t/2)$ and the probability of detecting $|e\rangle$ is $\cos^2(\Delta t/2)$ in the ideal case.}\label{ramseypicture}
\end{figure}

Ramsey spectroscopy is the measurement technique of transition frequency $\omega_0$ between the internal states of two-level atom \cite{Ramsey}. The system can only take one of two energy levels as they can denote the ground state $|g\rangle$ and the excited state $|e\rangle$. Standard Ramsey spectroscopy is operated as following. Initially, prepare $N$ ionized atoms confined in an ion trap \cite{wineland1998} and by optical pumping technique \cite{happer1970}, change the state of atoms into the same ground state, $|\psi_0\rangle=|g\rangle$. Then, the atoms are applied by $\pi/2$ pulse which leads each atom to be in the superposition state of the ground state and the excited state, $|\psi_1\rangle=(|g\rangle+|e\rangle)/\sqrt{2}$. After that, (classical) fields with frequency $\omega$ are applied to the atoms for interrogation time $t$ so that the state changes to $|\psi_2\rangle=(|g\rangle+e^{-i\Delta t}|e\rangle)/\sqrt{2}$ in a rotating frame, where $\Delta=\omega_0-\omega$ denotes the detuning between the atomic transition and classical driving field. Finally, the atoms are applied by the second $\pi/2$ pulse which changes the state to $|\psi_3\rangle=\sin(\Delta t/2)|g\rangle+\cos(\Delta t/2)|e\rangle$, and the internal state of each atom is measured by scattering light and detecting the fluorescence with a photomultipier (PMT). We estimate the transition frequency $\omega_0$ by counting the number of atoms in the excited state $|e\rangle$ and using the probability of detecting the excited state $|e\rangle$ in each atom which is given as
\begin{equation}\label{p1}
P(e|\omega_0)=|\langle e|\psi_3\rangle|^2=\cos^2\bigg(\frac{\Delta t}{2}\bigg)=\frac{1+\cos\Delta t}{2}.
\end{equation}
The probability of detecting the ground state is given as $P(g|\omega_0)=1-P(e|\omega_0)=\sin^2(\Delta t/2)$.
The operation of Ramsey spectroscopy is shown in Fig \ref{ramseypicture}.

From its functional form, the probability $P(e|\omega_0)$ can be used to estimate the frequency $\omega_0$ of the atomic state within a certain precision. The statistical fluctuation in the estimation which is associated with the given probability $P(e|\omega_0)$ can be obtained by using Cram\'{e}r-Rao inequality \cite{cramer,Helstrom},
\begin{equation}\label{cramer}
\delta\theta\geq\frac{1}{\sqrt{\nu F(\theta)}}
\end{equation}
where $F(\theta)=\sum_i \frac{1}{P(i|\theta)}\big(\frac{\partial P(i|\theta)}{\partial \theta}\big)^2$ is the Fisher information and $\nu$ is the number of repetition of trials. $P(i|\theta)$ is the conditional probability of obtaining a result $i$ with a given parameter $\theta$ and the summation is taken over all possible results. The Fisher information is a quantity that measures how much information can be obtained when the parameter $\theta$ is changed infinitesimally. It is known that the lower bound of the Cram\'{e}r-Rao inequality can be achieved asymptotically by the maximum likelihood estimator \cite{fisher}. The Fisher information plays an important role in estimation theory in that it gives a lower bound of statistical fluctuation.

In Ramsey spectroscopy, the unknown parameter to be estimated is the transition frequency $\omega_0$ between two internal states $|g\rangle$ and $|e\rangle$. Using the fact that possible results are the excited state $|e\rangle$ with the probability $P(e|\omega_0)$ and the ground state $|g\rangle$ with the probability $P(g|\omega_0)$, the Cram\'{e}r-Rao lower bound is easily obtained as
\begin{equation}\label{uncor}
|\delta \omega_0|=\frac{1}{\sqrt{\nu t^2}}=\frac{1}{\sqrt{NTt}},
\end{equation}
where $\nu=NT/t$ is the repetition of trials and $T$ is the total experimental time. Thus, the precision with $N$ atoms is proportional to $1/\sqrt{N}$, which is so-called the \textit{shot-noise limit} \cite{Itano 1993}. The shot-noise limit is originated from quantum projection noise. Note that the precision $|\delta \omega_0|$ is independent of the transition frequency $\omega_0$ with ignorance of decoherence effects.

So far, we have discussed in the ideal case. However, in a realistic experiment, it is necessary to include the effects of decoherence. As a realistic case, we introduce a model that was proposed in \cite{huelga1997}. In this model, dephasing of individual ions is the main decoherence, which is caused from collisions, stray fields, and laser instabilities. The effects of decoherence on a single ion $\rho$ can be described as the master equation in Lindblad form \cite{huelga1997,nielsen}:
\begin{equation}\label{decoherence}
\dot{\rho}(t)=i\Delta(\rho|e\rangle\langle e|-|e\rangle\langle e|\rho)+\frac{\gamma}{2}(\sigma_z\rho\sigma_z-\rho)
\end{equation}
where $\gamma$ is the decay rate and $\sigma_z=|g\rangle\langle g|-|e\rangle\langle e|$ denotes a Pauli spin operator. Eq. (\ref{decoherence}) is written in a rotating frame. The effect of decoherence is shown in the broaden signal (\ref{p1}) of a single ion:
\begin{equation}\label{p1d}
P_\gamma (e|\omega_0)=\frac{1+\cos(\Delta t) e^{-\gamma t}}{2},
\end{equation}
and again using Cram\'{e}r-Rao inequality (\ref{cramer}), the corresponding precision changes to
\begin{equation}\label{deco}
|\delta\omega_0^{\rm{dec}}|=\sqrt{\frac{1-\cos^2(\Delta t)e^{-2\gamma t}}{NTte^{-2\gamma t}\sin^2(\Delta t)}}.
\end{equation}
It can be found that as the decay rate $\gamma$ increases, the uncertainty becomes larger. In contrast to the ideal case, a precision in the presence of decoherence is dependent to the frequency $\omega_0$. When $\Delta t=k\pi/2$ $(k$ {\rm{is odd}}$)$ and $t=1/2\gamma\leq T$ are satisfied, minimum uncertainty is obtained as
\begin{equation}\label{minuncertainty}
|\delta \omega_0^{\rm{dec}}|_{\rm{min}}=\sqrt{\frac{2\gamma e}{NT}}.
\end{equation}
Note that minimum uncertainty is attained when the probabilities of obtaining the measurement outcomes $|g\rangle$ and $|e\rangle$ are same.

\subsection{Improvement of precision with GHZ correlation}
We have studied standard Ramsey spectroscopy using $N$ atoms, which allows us to obtain a shot-noise limit. It has been proposed that entanglement between atoms improves sensitivity of phase. Especially, one of the states that has been proposed to improve precision is maximally entangled multipartite state which is so-called Greenberger-Horne-Zeilinger (GHZ) state \cite{greenberger1990}
\begin{equation}
|\psi_{\rm{GHZ}}\rangle=\frac{|g\rangle^{\otimes N}+|e\rangle^{\otimes N}}{\sqrt{2}}
\end{equation}
where $|g\rangle^{\otimes N}\equiv|g\rangle |g\rangle \cdot\cdot\cdot |g\rangle$ and $|e\rangle^{\otimes N}\equiv|e\rangle |e\rangle \cdot\cdot\cdot |e\rangle$.
The GHZ states accumulate $N$ times amplified phase information than without entanglement, which results in a better precision. Implementation of GHZ states among atoms has been demonstrated by Cirac \textit{et. al.} \cite{cirac} and it will be explained in the following.

The preparation procedure of GHZ states is, in principle, that after all $N$ atoms are prepared in the ground state $|g\rangle^{\otimes N}=|g\rangle|g\rangle\cdot\cdot\cdot|g\rangle$, the first ion is applied by $\pi/2$ pulse to create the state $(|g\rangle+|e\rangle)|g\rangle|g\rangle\cdot\cdot\cdot|g\rangle/\sqrt{2}$ and the ions are operated by a "controlled-NOT(CNOT)" gate, the first ion as a controlled qubit and the second ion as target qubits to entangle the first two atoms, which changes the state to $(|g\rangle|g\rangle+|e\rangle|e\rangle)|g\rangle\cdot\cdot\cdot|g\rangle/\sqrt{2}$. Continuing the operation of CNOT gates, the first ion as controlled qubit and other ions as target qubits, the final state becomes the GHZ state $(|g\rangle|g\rangle\cdot\cdot\cdot|g\rangle+|e\rangle|e\rangle\cdot\cdot\cdot|e\rangle)/\sqrt{2}=(|g\rangle^{\otimes N}+|e\rangle^{\otimes N})/\sqrt{2}$. Despite theoretical straightforwardness of generating GHZ states, it is known that large size GHZ states are extremely difficult to create in practice \cite{sorensen2000,leibfried2003,leibfried2004,leibfried2005,monz2011}.

After preparation of GHZ states, (classical) fields of frequency $\omega$ are applied to the atoms for interrogation time $t$, which changes the state to $(|g\rangle^{\otimes N}+e^{-iN\Delta t}|e\rangle^{\otimes N})/\sqrt{2}$. Finally, the atoms are disentangled by the second set of controlled-NOT gates and the internal state of the first ion is measured. Again, we estimate the true frequency by the measurement outcomes and the probability of detecting $|e\rangle$,
\begin{equation}\label{pn}
P(e|N,\omega_0)=\frac{1+\cos N\Delta t}{2}=\cos^2\bigg(\frac{N\Delta t}{2}\bigg)
\end{equation}
which denotes the probability of detecting all $N$ atoms in the excited state. The probability of detecting all $N$ atoms in the ground state is given as $P(g|N,\omega_0)=1-P(e|N,\omega_0)=\sin^2(N\Delta t/2)$. The principle that GHZ states allow us to obtain a better precision is originated from the fact that the phase shift of GHZ states in interrogation time is amplified $N$ times than without entanglement in the same period, which results in the change of the probability of detecting $|e\rangle$.

As the previous single ion case, using the Cram\'{e}r-Rao inequality (\ref{cramer}), we attain a $\sqrt{N}$ improved precision than that of uncorrelated $N$ atoms (\ref{uncor}),
\begin{equation}\label{ghz}
|\delta\omega_0|=\frac{1}{\sqrt{\nu N^2 t^2}}=\frac{1}{N\sqrt{Tt}},
\end{equation}
where $\nu=T/t$ is the repetition of trials. Now, uncertainy of frequency is proportional to $1/N$, which is referred to as the Heisenberg limit. As the previous case, uncertainty does not depend on the frequency in the ideal case.

Taking into account decoherence, similarly to uncorrelated states, we can easily see that the signal (\ref{pn}) of GHZ states with $N$ ions becomes
\begin{equation}\label{pnd}
P_\gamma(e|N,\omega_0)=\frac{1+\cos(N\Delta t)e^{-N\gamma t}}{2},
\end{equation}
and uncertainty of frequency becomes
\begin{equation}\label{ghzdeco}
|\delta\omega_0^{\rm{dec}}|=\sqrt{\frac{1-\cos^2(N\Delta t)e^{-2N\gamma t}}{N^2 Tte^{-2N\gamma t}\sin^2(N\Delta t)}}.
\end{equation}
The exponential part in the denominator indicates that if $N$ is too large, uncertainty $|\delta \omega_0^{\rm{dec}}|$ goes to infinity, which means that a large size of GHZ state is fragile against decoherence. Again, uncertainty depends on the frequency and minimum uncertainty is achieved when $\Delta t=k\pi/2N$ $(k$ {\rm{is odd}}$)$ and $t=1/2\gamma N\leq T$ are satisfied. Minimum uncertainty is obtained when the probabilities of obtaining each result $|g\rangle$ and $|e\rangle$ are same. As a consequence, minimum uncertainty is same with that of uncorrelated states (\ref{minuncertainty}). Therefore, in the presence of decoherence GHZ states do not help attaining a better precision than that of uncorrelated states \cite{huelga1997}.

\subsection{Ambiguity of $\pi$-period in the frequency estimation}
\begin{figure}[t]
{\bf (a)} \hspace{7cm} {\bf (b)} \\
\begin{minipage}[h]{0.45\linewidth}
\includegraphics[scale =0.5]{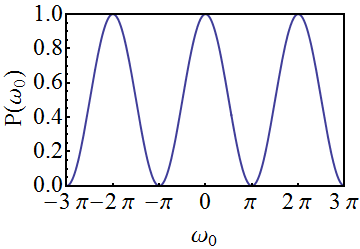}
\label{fig:combi}
\end{minipage}
\hspace{0.5cm}
\begin{minipage}[h]{0.45\linewidth}
\includegraphics[scale=0.5]{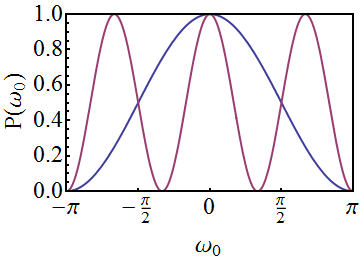}
\label{fig:combi}
\end{minipage}
\caption{Probability distribution of an excited state as a function of $\omega_0$ when $t=1$. (a) Periodicity of the probability distribution causes ambiguity in determining a single peak among the values in the periodic repetitions. If one knows that $\omega_0\in[-3\pi,3\pi)$, the estimate of $\omega_0$ can be taken from any values among $\pm2\pi$ and $0$. In order to avoid the ambiguity, we need to change the interaction time $t$ three times smaller than original. The new choice of interaction time allows us to discriminate the peak value of $\omega_0$ at $0$ from $\pm2\pi$. (b) Probability distibution of a single atom (blue) in an excited state and that of GHZ states with $3$ atoms (purple) are shown. If we know $\omega_0\in[-\pi,\pi)$ initially, the best choice of interrogation time is $t=1$. If we use a single atom, an estimation can be uniquely determined as $\widetilde{\omega_0}=0$. On the other hand, if we use GHZ states with $3$ atoms, the estimation becomes ambiguous because our estimation value is $\widetilde{\omega_0}=\pm 2\pi/3$ and $0$. To avoid the ambiguity, we need to choose $t$ three times smaller, which leads to the same uncertainty $|\delta \omega_0|$ with the scheme without GHZ correlation.}\label{periodicity}
\end{figure}

Let us consider a realistic experiment of frequency estimation with basic concepts of Ramsey spectroscopy. Basically, an estimation process is proceeded based on measurement outcomes which depend on the parameter to be estimated and the probability of obtaining the experimental data. In Ramsey spectroscopy, measurement outcomes are consisted of $|g\rangle$ and $|e\rangle$ only, and the probability of detecting the excited state $|e\rangle$ is given in (\ref{p1}) for a single atom. The probability (\ref{p1}) that we use in an estimation process has periodicity, which is shown in Fig \ref{periodicity}. A difficulty of frequency estimation comes from the periodicity because the periodicity prevents us from distinguishing a true estimate of frequency among a number of possible estimates. For example, let us suppose that the true frequency $\omega_0^{\rm{true}}$ is zero (we set $\omega=0$ for simplicity i.e. $\Delta=\omega_0$.), which leads to the measurement outcome $|e\rangle$ at all $N$ atoms. Then, our estimate obtained by the measurement outcomes is not $\omega_0=0$ but $\omega_0=2\pi m/t$ ($m$ an integer). Our estimation result is not unique. In other words, since two different frequencies $(\omega_0)_1=2\pi m/t$ and $(\omega_0)_2=2\pi n/t$ ($m$, $n$ integers) give the same experimental data, the data does not allow us to choose a single estimate of frequency. Hence, even after obtaining measurement outcomes, we are still required to determine a single estimation value. Determination of a unique estimate can be possible with the help of a prior knowledge about the true frequency $\omega_0$. If a provided prior knowledge about the true frequency is narrow enough to choose a unique estimate, the ambiguity problem is avoided. If a given prior knowledge is not enough narrow, we still have ambiguity in the estimation procedure. In short, because of periodicity, even when we try infinitely many measurements, it is not possible to determine a single estimation value without an appropriate prior knowledge. Thus, we are required to know about the true frequency initially with an enough degree of accuracy to choose a single estimation value so that we assume that an arbitrary prior knowledge about $\omega_0$ is always given in this paper.

In this situation, let us suppose that we are trying to improve precision. When we use $N$ atoms, the ultimate precision is given as (\ref{uncor}). Under a restriction of the limited number of atoms $N$ and a given total duration time $T$, a possible choice to improve preicision is increasing interrogation time $t$, since uncertainty of the frequency $|\delta \omega_0|$ decreases as the interrogation time $t$ increases. Hence, we can increase the time $t$ as possible in the only restriction $t\leq T$ where $T$ is sufficiently large. However, as the interrogation time $t$ increases, the period $2\pi/t$ of the signal (\ref{p1}) becomes shorter so that it causes ambiguity in determining a unique estimation value because a more number of frequencies become a possible estimation value as the period becomes shorter. In other words, the shorten period makes it harder to distinguish a true frequency from other possible estimates. Thus, while the longer interrogation time $t$ allows a better preicision, the increased interrogation time $t$ requires a narrower prior knowledge about the true frequency $\omega_0$ in order to determine a unique estimation value.

Let us suppose that we know that the true frequency $\omega_0$ lies in the interval $[0,\pi/L)$ where $L$ is a positive number. Under the prior knowledge, the largest possible value of interrogation time $t$ that allows us to determine a single peak is $L$. Then, the possible maximum interrogation time $t=L$ leads uncertainty to be
\begin{equation}\label{noprior}
|\delta\omega_0|=\frac{1}{\sqrt{NTL}}.
\end{equation}
This is minimum uncertainty with the given prior knowledge $\omega_0\in [0, \pi/L)$. It implies that minimum uncertainty of frequency estimation depends on the accuracy of a prior knowledge about the true frequency. In other words, if we know the true frequency more accurately at the beginning, we can obtain a better precision.

As previously discussed, quantum entanglement between atoms help us to improve precision of frequency estimation. Let us consider introducing quantum entanglement between atoms by exploiting GHZ states under the same prior knowledge that $\omega_0\in [0, \pi/L)$. Similar to the unentangled state case, the signal of GHZ states with $N$ atoms has a period $2\pi/Nt$ from (\ref{pn}). Therefore, the largest possible $t$ that allows us to choose a single estimation value is $L/N$. Eventually, uncertainty obtained by using GHZ states is same with (\ref{noprior}). The reason that uncertainty with uncorrelated state and with GHZ state is same is that the period of signal with GHZ states is $N$ times shorter than that of uncorrelated state so that we need to use a shorter interrogation time to avoid ambiguity in the estimation process. Consequently, in the consideration of a prior knowledge about the frequency, exploiting GHZ states does not improve precision, which is shown in Fig \ref{periodicity}. However, it does not mean that GHZ states are useless in improving the precision of frequency because it obviously has a better sensitivity locally. We introduce useful schemes that utilize the potential power of GHZ states and improve precision, avoiding ambiguity.

\section{Schemes to overcome ambiguity of $\pi$-period}
In this section, we analyze two measurement schemes to overcome ambiguity in frequency estimation of atomic states. Basic settings of the schemes are originally proposed by Pezze {\it et. al.} \cite{pezze2007} and Gkortsilas {\it et. al.}\cite{dunningham}. However, their schemes are limited either in the case of ideal measurement or in the case with photon losses. Here, we apply the schemes to frequency estimation with atomic GHZ states in the presence of decoherence. Decoherence effects on atomic states are different with photon losses in photonic states. Since photon losses can be recognized by comparing the number of detected photons with that of input photons, one is able to exclude measurement outcomes where photon losses occurred. Thus, one takes perfect measurement outcomes only in the estimation process. On the other hand, detection of internal states of atoms does not give any information whether decoherence had an effect on the atomic states. Therefore, all the measurement outcomes should be taken into account including atoms that were affected by decoherence. As such, decoherence in atomic states is different with photon losses.

The first scheme is using GHZ states with $1,2,2^2,...,2^{p-1}$ numbers of particles in order. It is known that this scheme allows phase estimation with the Heisenberg limit in the ideal case. The second scheme is using combinations of an uncorrelated state and GHZ states. The two schemes behave differently and give different precisions depending on the degree of decoherence effects. We analyze the two schemes in frequency estimation, considering decoherence effects on atomic states.

Before we introduce details of the schemes, let us review Bayesian analysis. Let us suppose that we know that the true frequency lies in $[0, \pi/L)$ again, and we choose $t=L$ which leads to the best sensitivity avoiding ambiguity for an uncorrelated state. According to the Bayes' theorem, the posterior probability distribution $P_p(\omega_0|N_T,X)$, given total resources $N_T$ and a data set of measurement outcomes $X=(x_1,x_2,...,x_{N_T})$, when $x_n\in \{g,e\}$, is given as
\begin{equation}\label{bayes}
P_p(\omega_0|N_T,X)=\frac{P(X|N_T,\omega_0)P(\omega_0)}{P(N_T,X)},
\end{equation}
where $P(\omega_0)$ is a prior distribution of $\omega_0$ and is completely flat for an unknown frequency, $P(\omega_0)=L/\pi$, and $P(X)$ is treated as a normalization constant. The posterior probability $P(\omega_0|N_T,X)$ is a conditional probability of frequency $\omega_0$ based on a data set $X$ of measurement outcomes obtained by using $N_T$ resources. We choose the maximum of the distribution as estimator $\widetilde{\omega_0}$ and uncertainty $|\delta \omega_0|$ given by $\int_{\widetilde{\omega_0}-\delta \omega_0}^{\widetilde{\omega_0}+\delta \omega_0} d\omega_0 P_p(\omega_0|N_T,X)=0.6827$ when $P_p(\omega_0|N_T,X)$ goes to normal distribution. Bayesian analysis allows us to obtain the estimation value and associated uncertainty by using measurement outcomes.

\subsection{GHZ states with particle number $1,2,4,...2^{p-1}$}

\begin{figure}[t]
\qquad \qquad \quad {\bf (a)} \\
\begin{minipage}[h]{0.67\linewidth}
\qquad \qquad \quad \includegraphics[scale =0.7]{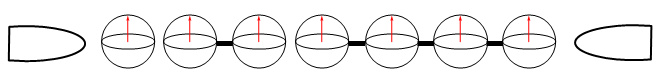}
\end{minipage} \\ \\
\qquad \qquad \quad {\bf (b)}
\center
\begin{minipage}[h]{0.5\linewidth}
\includegraphics[scale=0.55]{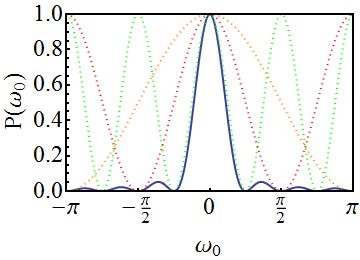}
\end{minipage}
\caption{Scheme with GHZ states with particle numbers $N_e=1,2,4...,2^{p-1}$ (a) Preparation of GHZ states with 1,2 and 4 atoms. The connection line between atoms denotes GHZ correlation. (b) Using GHZ states with 1,2 and 4 atoms, the overall probability distribution (blue solid line) has only a single peak. $\omega_0^{\rm{true}} t=0$. A prior knowledge about frequency is $\omega_0\in[-\pi,\pi)$. Orange, red and green dotted lines denote probability distributions of using GHZ states with 1, 2, and 4 atoms, respectively.}\label{GHZplot}
\end{figure}
Let us consider the first scheme : exploiting GHZ states with particle numbers $N_e=1,2,4,...,2^{p-1}$. Note that we have assumed that we know that $0\leq\omega_0^{\rm{true}}<\pi/L$ so that we have fixed the interrogation time $t=L$. Now let us suppose $\omega_0^{\rm{true}}=\pi/2L$ for simplicity. In this case, if we use a single atom, the probability of detecting the ground state and that of the excited state are same. Thus, the precision is maximum in the presence of decoherence. However, if we use GHZ states with even number of atoms, the measurement outcomes are either all $|g\rangle$ or all $|e\rangle$ where the precision is minimum for the realistic case, which is implied by (\ref{ghzdeco}). Thus, for GHZ states with the even number of atoms, we need to apply an additional phase shift $\pi/2$ in order to change the overall phase shift into one that gives minimum uncertainty where the probabilities of detecting $|g\rangle$ and $|e\rangle$ are same. Then, the effective frequency for GHZ states with the even number of atoms is $\omega_0+\pi/2N_e$.

Asymptotically, when we use GHZ states of $N$ atoms (If $N=1$, a single atom state) with the number of trials $\nu$ in frequency estimation, we obtain $\nu P(g|N,\omega_0^{\rm{eff}})$ number of the ground state and $\nu P(e|N,\omega_0^{\rm{eff}})$ number of the excited state. Due to the additional phase shift, the effective frequency $\omega_0^{\rm{eff}}$ is $\omega_0^{\rm{true}}+\pi/2N$ for GHZ states with even $N$ atoms and $\omega_0^{\rm{true}}$ for a single atom. Therefore, by using the Bayes' theorem (\ref{bayes}), when we use GHZ states of particle number $N_e=1,2,4,...,2^{p-1}$ with repetition $\nu=T/L$, the probability distribution in the ideal case becomes
\begin{eqnarray}\nonumber
P_p(\omega_0|N_T=\sum_{k=0}^{p-1}2^k,X)\propto\prod_{k=0}^{p-1} P(\omega_0|2^k,e)^{\nu P(e|2^k,\omega_0^{\rm{eff}})}\times P(\omega_0|2^k,g)^{\nu P(g|2^k,\omega_0^{\rm{eff}})}\\= [\cos^2(\omega_0t/2) \sin^2(\omega_0t/2)]^{\nu/2}\prod_{k=1}^{p-1}[\cos^2(2^k\omega_0 t/2-\pi/4)\sin^2(2^k\omega_0 t/2-\pi/4)]^{\nu/2}\nonumber \\ \simeq e^{-\nu(4^p-1)(\omega_0 t-\frac{\pi}{2})^2/6} \simeq e^{-\nu N_T^2 (\omega_0 t-\frac{\pi}{2})^2/6}
\end{eqnarray}
where $P(\omega_0|N_e,x)$ is defined by the Bayes' theorem (\ref{bayes}). We have used gaussian approximation $[\sin^2(\omega_0 t/2)]^{m}[\cos^2(\omega_0 t/2)]^{n}\sim e^{-\frac{m+n}{2}(\omega_0 t-\widetilde{\omega_0}t)^2}$ for large $m+n$, where $\widetilde{\omega_0}t=2\tan^{-1}(\sqrt{m/n})$ is the maximum point, and $N_T=\sum_{k=0}^{p-1} 2^k=2^p-1$. An advantage of using GHZ states with geometrically increasing number of atoms is cancellation of all peaks except the central one, which is shown in Fig \ref{GHZplot}. As a consequence, the probability distribution gives uncertainty of frequency which achieves the Heisenberg limit,
\begin{equation}
|\delta\omega_0|=\frac{\sqrt{3}}{\sqrt{\nu N_T^2L^2}}=\frac{\sqrt{3}}{N_T\sqrt{TL}}.
\end{equation}
In the realistic case, $P(x|N,\omega_0)$ are replaced by $P_\gamma(x|N,\omega_0)$. The results are shown in Fig \ref{optimization}. Since it achieves the Heisenberg limit  in the ideal case, apparently this scheme is helpful to utilize GHZ states under a prior knowledge and to improve precision. Nevertheless, this scheme is extremely fragile against decoherence because it exploits a large size of GHZ states. The result of the first scheme in the presence of decoherence indicates that it is required to use a number of uncorrelated atoms that are more robust against decoherence than GHZ states. Therefore, we need to introduce another scheme that is more robust against decoherence in a realistic situation.

\subsection{Combination of different correlations}
\begin{figure}[t]
\qquad {\bf (a)} \\
\begin{minipage}[h]{0.85\linewidth}
\qquad \qquad \includegraphics[scale =0.85]{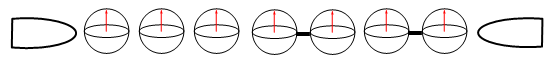}
\end{minipage} \\ \\
\qquad {\bf (b)} \hspace{7cm} {\bf (c)}
\center
\begin{minipage}[h]{0.45\linewidth}
\includegraphics[scale =0.5]{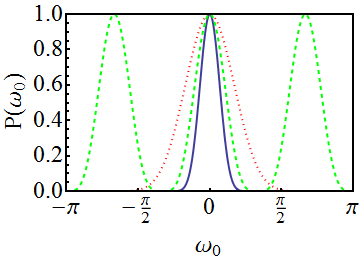}
\label{fig:combi}
\end{minipage}
\hspace{0.5cm}
\begin{minipage}[h]{0.45\linewidth}
\includegraphics[scale =0.5]{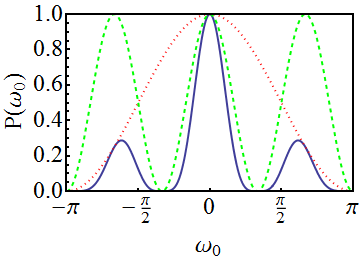}
\label{fig:combi2}
\end{minipage}
\caption{Scheme with combinations of uncorrelated atoms and GHZ states. $L=1$, $t=1$. (a) Preparation of a combination of three uncorrelated atoms and two copies of GHZ states with two atoms. The connection line between atoms denotes GHZ correlation. (b) An appropriate combination ($N_u=7, N_e=3$ and $p=4$) results in a single peak. (c) An inappropriate combination ($N_u=1, N_e=3$ and $p=2$) gives rise to a probability on the region that does not include a true value.  Green dashed line represents a probability distribution from $p$ copies of GHZ states with $N_e$ atoms and red dotted line represents a probability distribution from uncorrelated $N_u$ atoms. $\omega_0^{\rm{true}}=0$.}\label{schemes}
\end{figure}

The second scheme is to use appropriate combinations of uncorrelated atoms $N_u$ and the $p$ copies of GHZ states with $N_e$ atoms. Here, the uncorrelated atoms play a role in suppressing other periods except one period where the true frequency exists, while GHZ states play a role in improving sensivity, which is shown in Fig \ref{schemes}. Similar to the previous scheme, the probability distribution with repetition $\nu$ is asymptotically written as
\begin{eqnarray} \nonumber
P_p(\omega_0|N_T=N_u+p N_e,X)\propto \\ \qquad \qquad \qquad P(\omega_0|1, e)^{\nu N_u P(e|1,\omega_0^{\rm{true}})}\times P(\omega_0|1,g)^{\nu N_u P(g|1,\omega_0^{\rm{true}})}\nonumber
\\  \qquad \qquad \qquad \times P(\omega_0|N_e,e)^{\nu p P(e|N_e, \omega_0^{\rm{true}})}\times P(\omega_0|N_e,g)^{\nu p P(g|N_e,\omega_0^{\rm{true}})}.\label{asymp}
\end{eqnarray}
To minimize the standard deviation of the probability distribution, we need to choose optimal $N_u,N_e$ and $p$ numerically. For numerical optimization, first of all, we obtain the probability distribution in (\ref{asymp}) by using (\ref{p1}) and (\ref{pn}) in ideal case or by using (\ref{p1d}) and (\ref{pnd}) in realistic case. Then, for a proper value of repetition $\nu$ and fixed $N_T=N_u+p N_e$, by changing $N_u,N_e$ and $p$ appropriately, we find values of $N_u,N_e$ and $p$ that minimize the standard deviation of the probability distribution. Then, we iterate the same procedure for different values of $N_T$. After the numerical optimization, only one peak survives where the true frequency exists.

\begin{figure}[t]
{\bf (a)  \hspace{7cm} \bf (b)} \\
\begin{minipage}[h]{0.45\linewidth}
\includegraphics[scale =0.5]{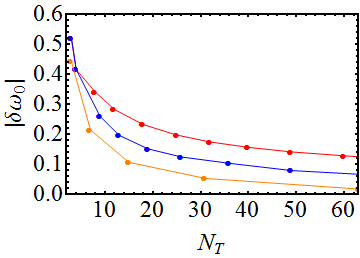}
\label{fig:noloss1}
\end{minipage}
\hspace{0.5cm}
\begin{minipage}[h]{0.45\linewidth}
\includegraphics[scale =0.5]{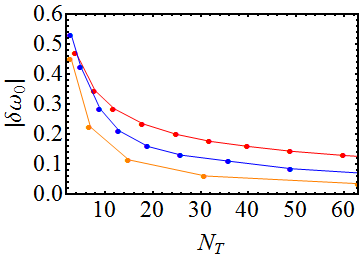}
\label{fig:loss1}
\end{minipage} \\
{\bf (c) \hspace{7cm} \bf (d)} \\
\begin{minipage}[h]{0.45\linewidth}
\includegraphics[scale =0.5]{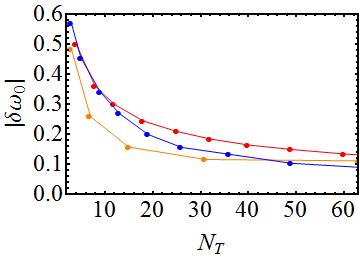}
\label{fig:noloss1}
\end{minipage}
\hspace{0.5cm}
\begin{minipage}[h]{0.45\linewidth}
\includegraphics[scale =0.5]{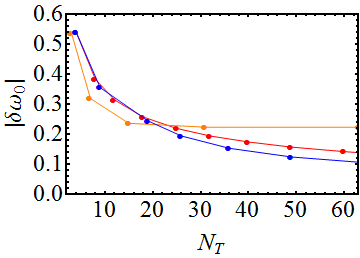}
\label{fig:loss1}
\end{minipage}
\caption{The comparison of precision with various decoherence rates $\gamma$. We fix $L=1$, $t=1$ and $\omega_0^{\rm{true}}=\pi/2$. (a) In the absence of decoherence (ideal case), exploiting GHZ states with $1,2,4,...,2^{p-1}$ atoms (orange) achieves the Heisenberg limit and the uncorrelated states (red) attain shot-noise limit and combinations of uncorrelated atoms and GHZ states(blue) achieves the sub shot-noise limit. (b)(c)(d) As the decoherence $\gamma$ increases, the scheme with GHZ states with $1,2,4,...,2^{p-1}$ atoms losses its advantage since GHZ states with a number of atoms are fragile against decoherence. However, the scheme with combinations of uncorrelated states and GHZ states beats the shot-noise limit even in the presence of decoherence. (b) $\gamma=0.01$. (c) $\gamma=0.05$. (d) $\gamma=0.1$.}
\label{optimization}
\end{figure}

Let us consider the same situation with the first scheme, a prior knowledge that $\omega_0^{\rm{true}}\in[0,\pi/L)$, fixed $t=L$ and the true frequency $\omega_0^{\rm{true}}=\pi/2L$. In this case, we only use GHZ states with odd number of atoms because GHZ states with odd number of atoms attain minimum uncertainty at $\omega_0^{\rm{true}}=\pi/2L$ in the presence of decoherence. At $\omega_0^{\rm{true}}=\pi/2L$, the probability of detecting $|g\rangle$ and that of detecting $|e\rangle$ are $1/2$ for both uncorrelated states and GHZ states with odd number of atoms. Substituting $\omega_0^{\rm{true}}=\pi/2L$ into (\ref{asymp}), in the ideal case, the asymptotic probability distribution with the combination of $N_u$ uncorrelated atoms and $p$ copies of GHZ states of $N_e$ atoms and repetition $\nu=T/t$ can be obtained as
\begin{eqnarray}
P_p(\omega_0|N_T=N_u+p N_e,X)\nonumber \\
=[\cos^2(\omega_0 t/2)\sin^2(\omega_0 t/2)]^{\nu N_u/2}[\cos^2(N_e \omega_0 t/2)\sin^2(N_e \omega_0 t/2)]^{\nu p/2}\nonumber \\ \simeq C e^{-(N_u+p N_e^2)(\omega_0 t-\pi/2)^2/2}
\end{eqnarray}
where $C$ is a normalization constant. The gaussian approximation is valid when the probability distribution is well-localized around $\pi/2$. If $N_u, N_e$ and $p$ are numerically optimized so that the probability distribution has a single peak, the approximation is valid. After the numerical optimization, uncertainty of frequency is obtained as
\begin{equation}
|\delta \omega_0|=\sqrt{\frac{1}{\nu (N_u+p N_e^2)L^2}}=\sqrt{\frac{1}{(N_u+p N_e^2)TL}}
\end{equation}
which is the consistent with the Cram\'{e}r-Rao bound. It implies that the Cram\'{e}r-Rao bound is saturated by Bayesian analysis if there is no ambiguity in the estimation process. Uncertainty obtained by the numerical optimization of $N_u, N_e$ and $p$ is shown in Fig \ref{optimization}. In the realistic case, $P(x|N,\omega_0)$ are replaced by $P_\gamma(x|N,\omega_0)$ in  (\ref{asymp}).

In the absence of decoherence, exploiting the GHZ states with $1,2,4,...,2^{p-1}$ atoms allows us to achieve the Heisenberg limit as known. A drawback of this scheme is that  as the decoherence rate $\gamma$ increases, uncertainty becomes larger as shown in Fig \ref{optimization}. Especially, at $\gamma=0.1$ for large $N_T$, uncertainty is larger than that of uncorrelated states. In other words, since a large size of GHZ state is fragile against decoherence, this scheme is not practical when the decoherence rate is large. On the contrary, when we use appropriate combinations of an uncorrelated state and GHZ states, since we use a number of uncorrelated atoms, we can expect the robustness against decoherence. Indeed, Fig \ref{optimization} shows that the scheme with combinations of an uncorrelated state and GHZ states is robust against decoherence. More generally, it is noticed that the best scheme depends on decoherence rate $\gamma$. In the ideal case, the first scheme attains Heisenberg limit, which is the best sensitivity among our schemes. As the decoherence rate $\gamma$ increases, however, the second scheme becomes better. Thus, depending on the decoherence rate $\gamma$, we need to choose an optimal scheme.

In the second scheme, we have used combinations of uncorrelated atoms and GHZ states with fixed number of atoms for simplicity. However, it is possible to use combinations of uncorrelated states and various sizes of GHZ states. Indeed, the first scheme is one of the combinations of various GHZ states and the first scheme gives a better precision when the decoherence rate $\gamma$ is small. Therefore, it can be considered to use different various combinations depending on the decoherence rate $\gamma$.

We have assumed that the true frequency $\omega_0$ is $\pi/2L$ in the both schemes because the best sensitivity is attained at $\pi/2L$ for uncorrelated atoms in the realistic case. If the phase shift $\omega_0 L$ is not $\pi/2$, we need to modify the phase shift to be $\pi/2$ in order to attain minimum uncertainty, which is realizable by a feedback mechanism \cite{higgins, dunningham}. A feedback mechanism is implemented by the Bayesian analysis on the collected data. After each trial, we apply an additional phase shift which is determined by the Bayesian analysis to make the total phase shift to be $\pi/2$. For every trial, we repeat this process and then total phase shift becomes $\pi/2$ which gives minimum uncertainty of frequency.

One of the advantages in the second scheme is that it does not require a large size of GHZ state which is experimentally difficult to generate and has a short coherence time \cite{sorensen2000,leibfried2003,leibfried2004,leibfried2005,monz2011}. Indeed, the largest GHZ states that we have used in the numerical optimization are those of 5 atoms in both the ideal case and the realistic case. It indicates that our scheme is practical as well as useful. Nevertheless, since generating GHZ states requires extremely delicate experiment and the coherence time is very short, realization of our scheme may necessitate more advanced experiment devices and skills than current.

\section{Conclusion}
Periodicity of probability distribution causes ambiguity in the frequency estimation process. Furthermore, in the consideration of a prior knowledge, exploiting GHZ states does not help improving precision of frequency estimation when one uses only GHZ states with the same number of atoms. In order to avoid the ambiguity and utilize GHZ states for improvement of precision, we implement two different schemes. The first scheme is employing GHZ states with $1,2,4,...,2^{p-1}$ atoms, which improves precision significantly in the ideal case. Nevertheless, since a large size of GHZ states is fragile against decoherence, the first scheme is no longer advantageous in the realistic case. The second scheme that is robust against decoherence is exploiting appropriate combinations of uncorrelated atoms and GHZ states. As the decoherence rate $\gamma$ increases, it is shown that the sensitivity of second scheme can be better than that of the first scheme. In addition, we conclude that the scheme with combinations of uncorrelated and GHZ states allows us to achieve a sub shot-noise precision in the presence of decoherence.

\section*{Acknowledgement}
This work was supported by the ICT R\&D program of MSIP/IITP
(No.2014-044-014-002) and the National Research Foundation of Korea (NRF) grant funded by the Korean Government (No.NRF-2013R1A1A2010537).

\end{document}